\documentclass[10pt,aip,reprint]{revtex4-1}
\usepackage{amsmath,amsfonts,
	amsgen,amsbsy,amsopn,amstext,amssymb
}
\usepackage{booktabs}
\usepackage{graphicx}% Include figure files
\usepackage{dcolumn}
\newcolumntype{.}{D{.}{.}{-1}}
\usepackage{epstopdf}
\usepackage{color}
\renewcommand{\i}{i}
\renewcommand{\(}{\left(}
\renewcommand{\)}{\right)}
\renewcommand{\Pr}{\mathrm{Pr}}

\begin{document}
\title{Rainbow-trapping absorbers: Broadband, perfect and asymmetric sound absorption by subwavelength panels with ventilation}

\author{No\'e Jim\'enez*}
\author{Vicent Romero-Garc{\'i}a}
\author{ Vincent Pagneux}
\author{Jean-Philippe Groby}
\affiliation{Laboratoire d'Acoustique de l'Universit\'e du Maine - CNRS UMR 6613, Le Mans, France} %\altaffiliation[Also at ]{Physics Department, XYZ University.}%Lines break
%\email{noe.jimenez@univ-lemans.fr}
%\affil[+]{these authors contributed equally to this work}


\begin{abstract}
Perfect, broadband and asymmetric sound absorption is theoretically, numerically and experimentally reported by using subwavelength thickness panels in a transmission problem. The panels are composed of a periodic array of varying cross-section waveguides, each of them being loaded by an array of Helmholtz resonators (HRs) with graded dimensions. The low cut-off frequency of the absorption band is fixed by the resonance frequency of the deepest HR, that reduces drastically the transmission. The preceding HR is designed with a slightly higher resonance frequency with a geometry that allows the impedance matching to the surrounding medium. Therefore, reflection vanishes and the structure is critically coupled. This results in perfect sound absorption at a single frequency. We report perfect absorption at 300 Hz for a structure whose thickness is 40 times smaller than the wavelength. Moreover, this process is repeated by adding HRs to the waveguide, each of them with a higher resonance frequency than the preceding one. Using this frequency cascade effect, we report quasi-perfect sound absorption over almost two frequency octaves ranging from 300 to 1000 Hz for a panel composed of 9 resonators with a total thickness of 11 cm, i.e., 10 times smaller than the wavelength at 300 Hz.
\end{abstract}

%\flushbottom
\maketitle

\section{Introduction}\label{sec:intro}
%\linenumbers
Wave manipulation using metamaterials has been extensively studied in electromagnetism \cite{zheludev2012}, elasticity \cite{ding2007, christensen2015} or acoustics \cite{yang2008, cummer2016}, and among the very innovative systems that have been demonstrated are the metamaterial wave absorbers \cite{landy2008,watts2012,cui2014,lee2016,cui2012,ding2012}. In the particular case of sound waves, the selective bandwidth of most studied metamaterials limits their practical applications for audible frequencies: the audible frequency band covers more than ten frequency octaves, while in contrast, visible light spectrum covers less than one octave. Nevertheless, acoustic metamaterials have found practical applications in the design of selective low-frequency sound absorbing materials composed of membrane-type resonators \cite{yang2008,mei2012,ma2014,romero2016}, quarter-wavelength resonators (QWRs) \cite{jiang2014,leclaire2015,groby2015,groby2016,li2016} and Helmholtz resonators (HRs) \cite{romero2016use,jimenezAPL2016,Jimenez2017,achilleos2017}.

By using the strong dispersion produced by local resonances, slow sound can be generated inside acoustic materials \cite{santillan2011}. Recently, slow sound phenomena have been exploited to design deep-subwavelength thickness absorbing structures \cite{leclaire2015,groby2015,groby2016,jimenezAPL2016,Jimenez2017}. Of particular interest are perfect absorbing materials, facing the challenge of impedance mismatch to the surrounding medium. To achieve perfect absorption, the intrinsic losses of the system must exactly compensate the energy leakage at one resonance of the structure \cite{chong2010,wan2011,romero2016,romero2016use}. When this condition is fulfilled the system is critically coupled with the exterior medium and perfect absorption is observed. 

Perfect acoustic absorption has been reported in rigidly-backed subwavelength structures by using slow sound and QWRs \cite{groby2016} or HRs \cite{jimenezAPL2016}, or by using membranes and plates \cite{yang2008,ma2014,romero2016}. Until now, only a few works have presented perfect and broadband absorption in rigid-backed subwavelength metamaterials. One kind of broadband absorbers are metaporous materials \cite{groby2011,lagarrigue2013,boutin2013,groby2015b}, whose low frequency performance is limited by the inertial regime of the porous matrix material. Other configurations include the use of panels composed of a parallel arrangement of different QWRs designed to be impedance matched at selected frequencies, e.g, the used in optimized absorbers based in sound diffusers \cite{wu2000} or, in the same way, optimally designed panels to the limit imposed by causality \cite{yang2017}. The use of poroelastic plates that exhibit low quality factor resonances to extend the absorption bandwidth has been also proposed \cite{romero2016}. In Ref~\cite{jiang2014} the authors used a graded set of QWRs in a slightly-subwavelength thickness structure to obtain quasi-perfect and broadband absorption. This last configuration can also fulfil the critical coupling conditions at more than one frequency, then, exhibiting perfect and broadband absorption \cite{groby2016}. In Ref.~\cite{romero2016use} a similar approach was presented using detuned HRs in a rigidly-backed waveguide. Finally, an extension of these ideas has been used to produce multiple slow waves inside a rigidly-backed graded structure of porous material to improve the broadband behaviour \cite{yang2016}, but critical coupling conditions were not fulfilled at most resonances and perfect absorption was only observed at a single frequency.

However, when the system is not rigidly-backed and transmission is allowed, obtaining perfect absorption becomes challenging because the scattering matrix of the system presents two different eigenvalues. In order to obtain perfect absorption both eigenvalues must vanish at the same frequency \cite{merkel2015}. This implies that symmetric and antisymmetric modes must be simultaneously critically coupled at a given frequency \cite{piper2014}. When the eigenvalues are both zero but at different frequencies, then the system cannot present perfect absorption, but quasi-perfect absorption can be achieved by approaching the symmetric and antisymmetric modes using strong dispersion \cite{Jimenez2017}. Perfect acoustic absorption in transmission problems can be obtained by using degenerate resonators, exciting a monopolar and a dipolar mode at the same frequency \cite{yang2015}. Using elastic membranes decorated with designed patterns of rigid platelets \cite{mei2012} very selective low-frequency perfect absorption can be observed. Another strategy consists in using asymmetric graded materials, e.g., chirped layered porous structures \cite{jimenez2016broadband}, but these structures lack of subwavelength resonances and therefore its thickness is of the order of half of the incoming wavelength. A final configuration to achieve perfect absorption in transmission consists in breaking the symmetry of the structure by making use of double-interacting resonators, then perfect absorption was observed in waveguides at a particular frequency \cite{merkel2015}.

\begin{figure*}[t]
	\centering
	\includegraphics[width=0.9\textwidth]{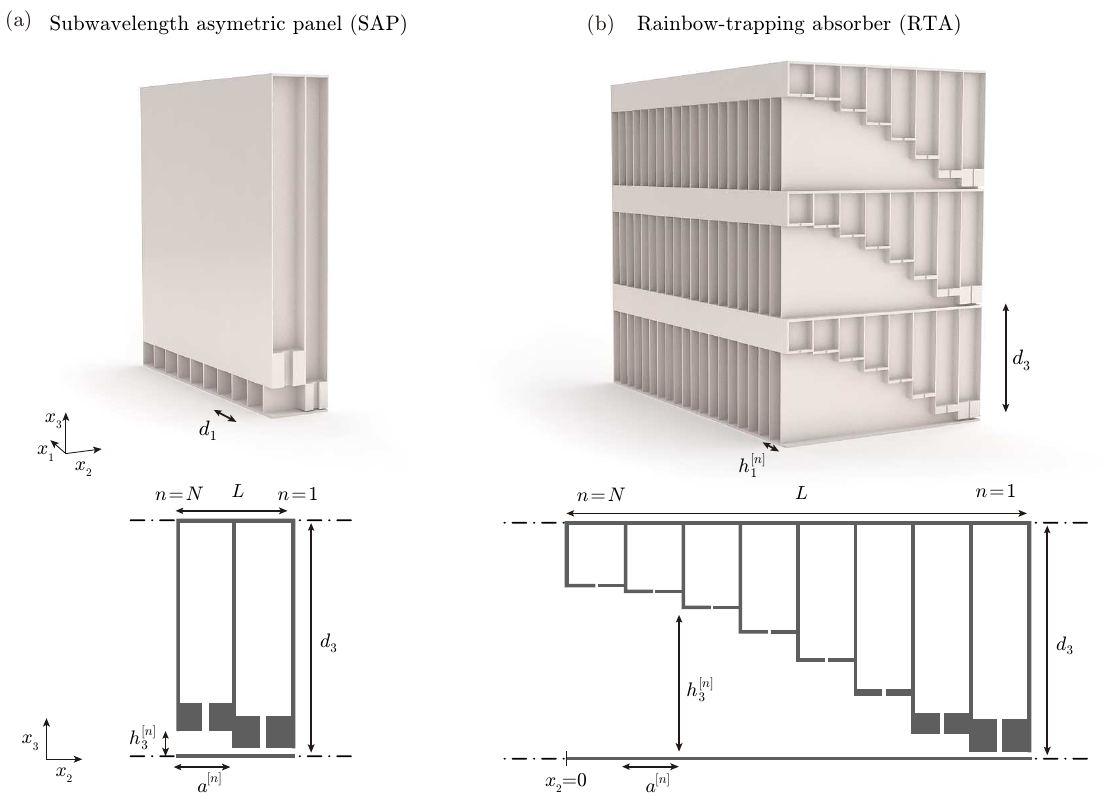}
	\caption{(a) Conceptual view of a subwavelength asymmetric panel (SAP) ($N=2$ resonators), where cross-section shows the waveguide and the loading HRs. (b) Conceptual view of a rainbow trapping absorber (RTA) with $N=8$ HRs. Scheme showing the geometrical variables for (c) the SAP and (d) RTA panels.}
	\label{fig:Fig1}
\end{figure*}

In this work, we address the problem of perfect and broadband acoustic absorption using deep-subwavelength structures, which to our knowledge was never addressed before in non rigidly-backed panels. To do so, we design panels composed of monopolar resonators with graded dimensions, namely \textit{rainbow-trapping absorbers}. The designed panels present broadband, perfect and asymmetric sound absorption, and, due to slow sound, their thickness is reduced to the deep-subwavelength regime. Rainbow trapping phenomenon, i.e., the localization of energy due to a gradual reduction of the group velocity in graded structures, has been observed in optics \cite{tsakmakidis2007trapped}, acoustics \cite{zhu2013acoustic,romero2013} or elastodynamics \cite{colombi2016}. However, losses were not accounted for and, therefore, absorption was not studied in these works. In the present configuration, a set of graded HRs is used, allowing to reduce, in addition to the thickness of the panels, the dimension of the unit cell to the deep-subwavelength regime. Using QWRs \cite{jiang2014,yang2016}, rigidly-backed absorbers are about 4 times smaller than the wavelength in the transverse direction of propagation, while in the configuration presented in this paper, and using non-rigidly backed conditions, the size of the structure in the transverse direction is up to 30 times smaller than the absorbed wavelength. 

In particular, the structures are composed of a rigid panel, of thickness $L$, periodically perforated with series of identical waveguides of variable square cross-section loaded by an array of $N$ HRs of different dimensions, as shown in Figs.~\ref{fig:Fig1}~(a,~b). Each waveguide is therefore divided in $N$ segments of length $a^{[n]}$, width $h_1^{[n]}$ and height $h_3^{[n]}$. The HRs are located in the middle of each waveguide section. Two samples were designed. The first one, namely \textit{subwavelength asymmetric panel} (SAP)  was composed of $N=2$ HRs and it is shown in Fig.~\ref{fig:Fig1}~(a). The second one, namely \textit{rainbow-trapping absorber} (RTA) was composed first, in a design stage, by $N=8$ and, finally, $N=9$ HRs for the experimental tests, as shown in Fig.~\ref{fig:Fig1}~(b). The SAP was designed to produce a single-frequency peak of perfect absorption while the RTAs were designed to exhibit broadband perfect absorption. The geometrical parameters of both structures were tuned using optimization methods (sequential quadratic programming (SQP) \cite{powell1978}). In the case of the SAP ($N=2$) the cost function minimized during the optimization process was $\varepsilon_{\rm SAP}=|R^-|^2+|T|^2$, i.e., to maximize the absorption at a given frequency, in this case we selected 300 Hz. The length of the SAP was constrained to $L=2.64$ cm, i.e., a panel 40 times thinner than the incoming wavelength. In the case of the rainbow trapping absorber ($N=9$), the cost function was $\varepsilon_{\rm RTA}=\int_{f_1}^{f_N}|R^-|^2+|T|^2 df$, i.e., to maximize the absorption in a broad frequency bandwidth, that was chosen from $f_1=300$ to $f_N=1000$ Hz. In the case of the RTA the length of the panel was constrained to $L=11.3$ cm, i.e. a panel 10 times thinner than the wavelength at 300 Hz. The geometrical parameters obtained by the optimization process are given in Tables \ref{table:tableSAP2},\ref{table:tableRTA9} for the SAP and RTA, respectively.

% ============================================================================
\section{Results}\label{sec:results}
\subsection{Subwavelength asymmetric panel (SAP)}\label{sec:asymmetric}
%\subsubsection*{Monochromatic perfect absorption}
\begin{figure*}[t]
	\centering
	\includegraphics[width=0.8\textwidth]{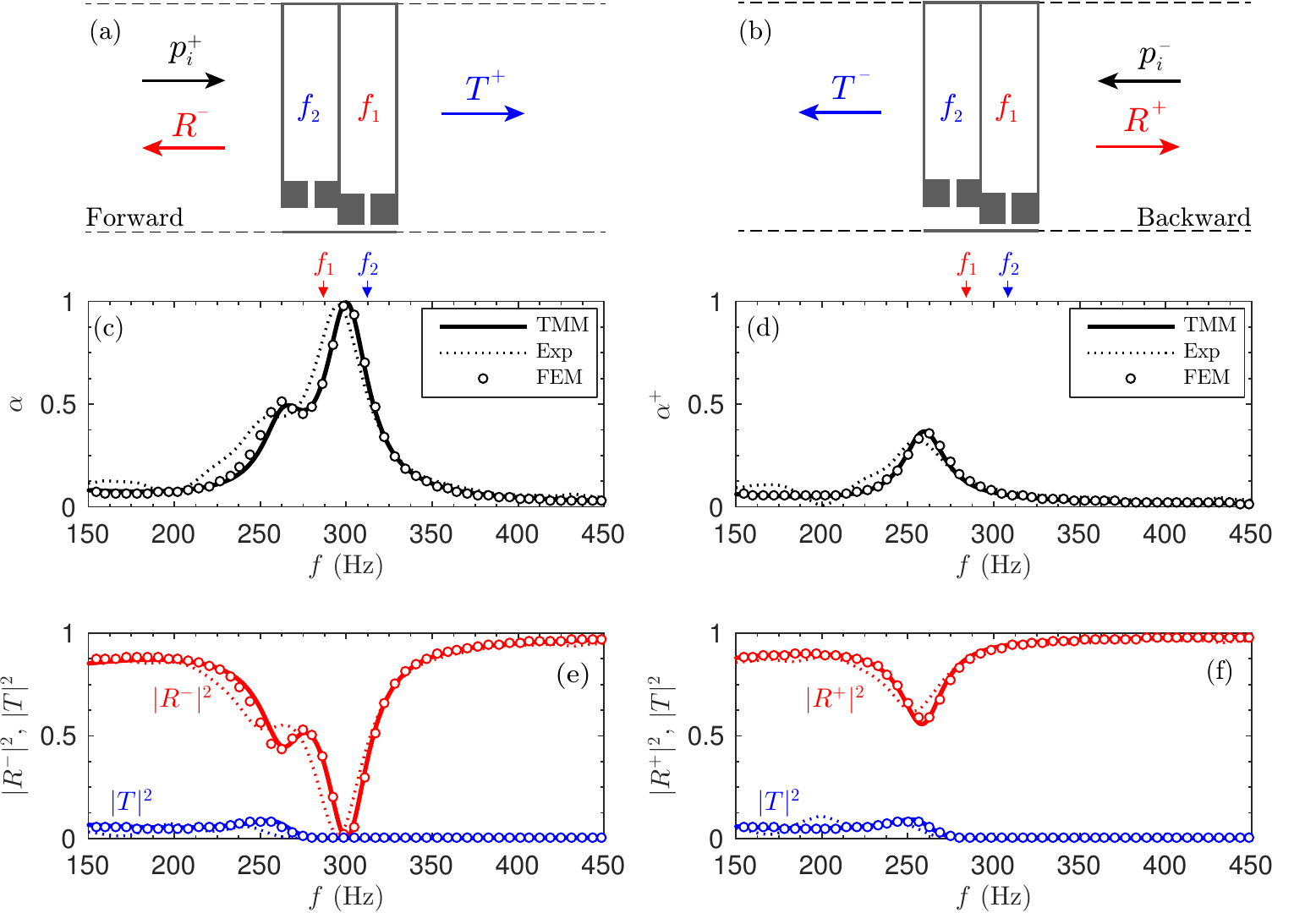}
	\caption{Scheme of the subwavelength asymmetric panel in (a) forward and (b) reverse configuration. (c) Absorption for the forward configuration obtained using TMM (continuous line), FEM (circles), and experiment (dotted line). Corresponding reflection and transmission coefficients. (d) Absorption for the backward configuration. (f) Corresponding reflection and transmission coefficients. The arrows mark the resonance frequencies of the HRs, $f_1$ and $f_2$.}
	\label{fig:Fig2}
\end{figure*}
We start analysing the behaviour of the designed SAP with $N=2$ HRs, considering the two directions of incidence, namely \textit{forward} and \textit{backward}, as depicted in Figs.~\ref{fig:Fig2}~(a,b). Figures~\ref{fig:Fig2}~(c-f) show the corresponding absorption, reflection and transmission coefficients for each case. The results is calculated analytically using the transfer matrix method (TMM) in which the thermoviscous losses are accounted for, numerically using finite element method (FEM) and experimentally validated using stereo-lithographic 3D printed structures and impedance tube measurements. See Section Methods for further details. Good agreement is observed between analytical, numerical and experimental results. 

First, in the forward configuration, shown in Fig.~\ref{fig:Fig2}~(a), the resonator $n=1$ of the waveguide presents a resonance frequency at $f_1=285$ Hz. As a consequence, above $f_1$, a band gap is introduced and the transmission is strongly reduced, the HR acting effectively as a rigidly-backed wall for the right ingoing waves. Then, the resonator $n=2$, with a superior resonance frequency at $f_2=310$ Hz, is tuned by the optimization process to critically couple the system with the exterior medium, matching the impedance of the waveguide to that of the surrounding medium. This is achieved at 300 Hz. As a consequence, no reflected waves are produced at this particular frequency and therefore, $\alpha=1-|R^-|^2-|T|^2 = 1$ holds. In this situation, perfect absorption is observed in a panel with a thickness 40 times smaller that the wavelength, i.e., a panel of thickness $L=2.64$ cm. It is worth noting here that the change of section in the main waveguide helps to achieve the impedance matching, specially for very thin SAPs as the one presented here. We will see later on that this steeped change in the cross-section is analogous to the graded profile of the main waveguide for the broadband structure. 

Second, in the backward propagation shown in Fig.~\ref{fig:Fig2}~(b), the wave impinges first the lowest resonance frequency resonator, $f_1$. Now at 300 Hz the wave almost no transmission is allowed in the waveguide. As the waveguide is not impedance matched at 300 Hz in backward direction, reflection is high and absorption is poor ($\alpha^+=0.05$). For frequencies below $f_2$, propagation is allowed in the main waveguide and the effect of the second HR may be visible inducing a decrease of the reflection coefficient. However, the impedance matching in the backward direction is not fully achieved and only a small amount of absorption is observed near the resonance frequency of the first resonator. Therefore, the absorption in this configuration is highly asymmetric. 

\begin{figure*}[tbp]
	\centering
	\includegraphics[width=0.8\textwidth]{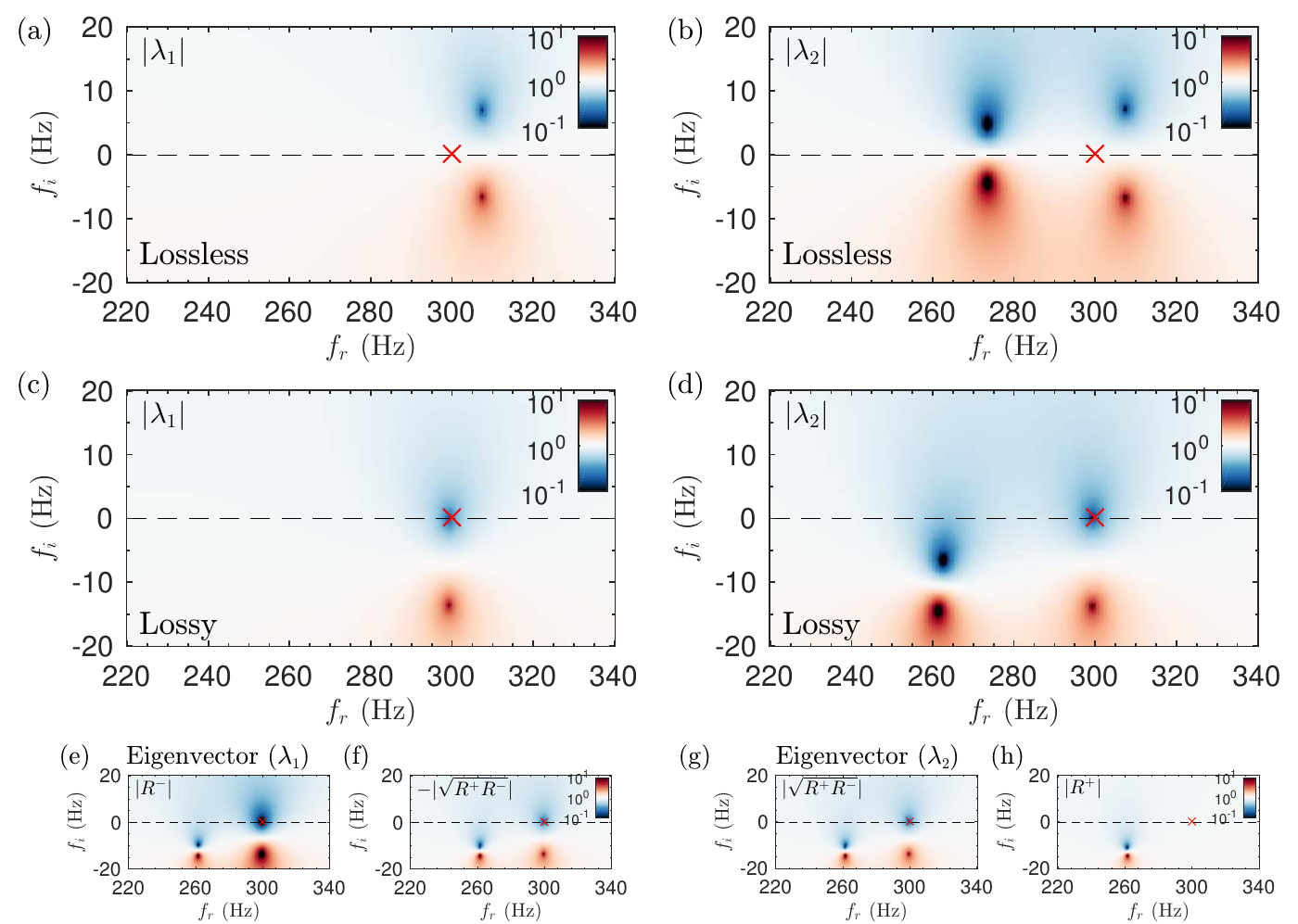}
	\caption{Complex frequency representation of the two eigenvalues of the scattering matrix, $|\lambda_1|$, $|\lambda_2|$, (a-b) in the lossless case and (c-d) including the thermo-viscous losses in the resonators and the main waveguides. Colormap in $10 \log_{10}|\lambda|^2$ scale.}
	\label{fig:Fig3}
\end{figure*}

%\subsubsection*{Complex frequency plane analysis}
In order to go further in the physical understanding of the scattering problem, we analyse the eigenvalues and eigenvectors of the \textbf{S}-matrix of the SAP in the complex frequency plane. At this stage, it is worth noting that perfect absorption can only be obtained when the two eigenvalues of the \textbf{S}-matrix are zero at the same purely real frequency. Let us start by considering the lossless propagation, neglecting the thermo-viscous losses by using the air parameters in the calculations. It can be observed that the eigenvalues shown in Fig.~\ref{fig:Fig3}~(a,~b) present pairs of zeros and poles, each pole identified with a resonance of the system. The position of the zeros and poles in the complex frequency plane characterizes the physical transmission and reflection properties along the real frequency axis. In the lossless case each pole presents a zero that is its complex conjugate, i.e., they are located at the same real frequency but in the opposite half-space\cite{romero2016use}.

Now we turn to the lossy case, accounting for the visco-thermal losses in the resonators and the waveguides. In general, once losses are introduced the zero-pole structure of a given system is translated in the complex plane (also slightly deformed), approaching the zeros to the real frequency axis. Figures~\ref{fig:Fig3}~(c,~d) show the eigenvalues of the \textbf{S}-matrix once thermo-viscous losses are introduced. At 300 Hz, both eigenvalues, $\lambda_{1,2}$, given by Eq.~(\ref{eq:lambda}), present a zero exactly located at the real axis. This implies $T=\sqrt{R^+R^-}=0$. When this occurs exactly at the real frequency axis and at the same frequency, it implies perfect absorption, as it was observed in the previous section. 

However, the information provided by the eigenvalues is not sufficient to determine from which side the perfect absorption is produced. The only information provided when the eigenvalues are zero is that the transmission coefficient and at least one of the two reflection coefficients vanish at this frequency. Thus, in order to draw the complete picture of the physical process, the corresponding eigenvectors are analysed. Figures~\ref{fig:Fig3}~(e,~f) and Figs.~\ref{fig:Fig3}~(g,~h) show the eigenvectors corresponding to $\lambda_1$ and $\lambda_2$, given by Eqs.~(\ref{eq:v1}), respectively. From the complex frequency representation of the eigenvectors we can obtain the remaining information, i.e., the reflection coefficients from each side. In this case, we can see that $R^-=0$ at 300 Hz, while $R^+\neq0$. This implies that the structure is only critically coupled with the surrounding medium when the incoming wave impinges the SAP from one side, i.e., in the forward direction the present case. Therefore, asymmetric perfect absorption is observed. In particular, $\alpha=1$ and $\alpha^+<1$ for the designed SAP at the critical coupling frequency ($f=300$ Hz). 

\begin{figure*}[t]
	\centering
	\includegraphics[width=\textwidth]{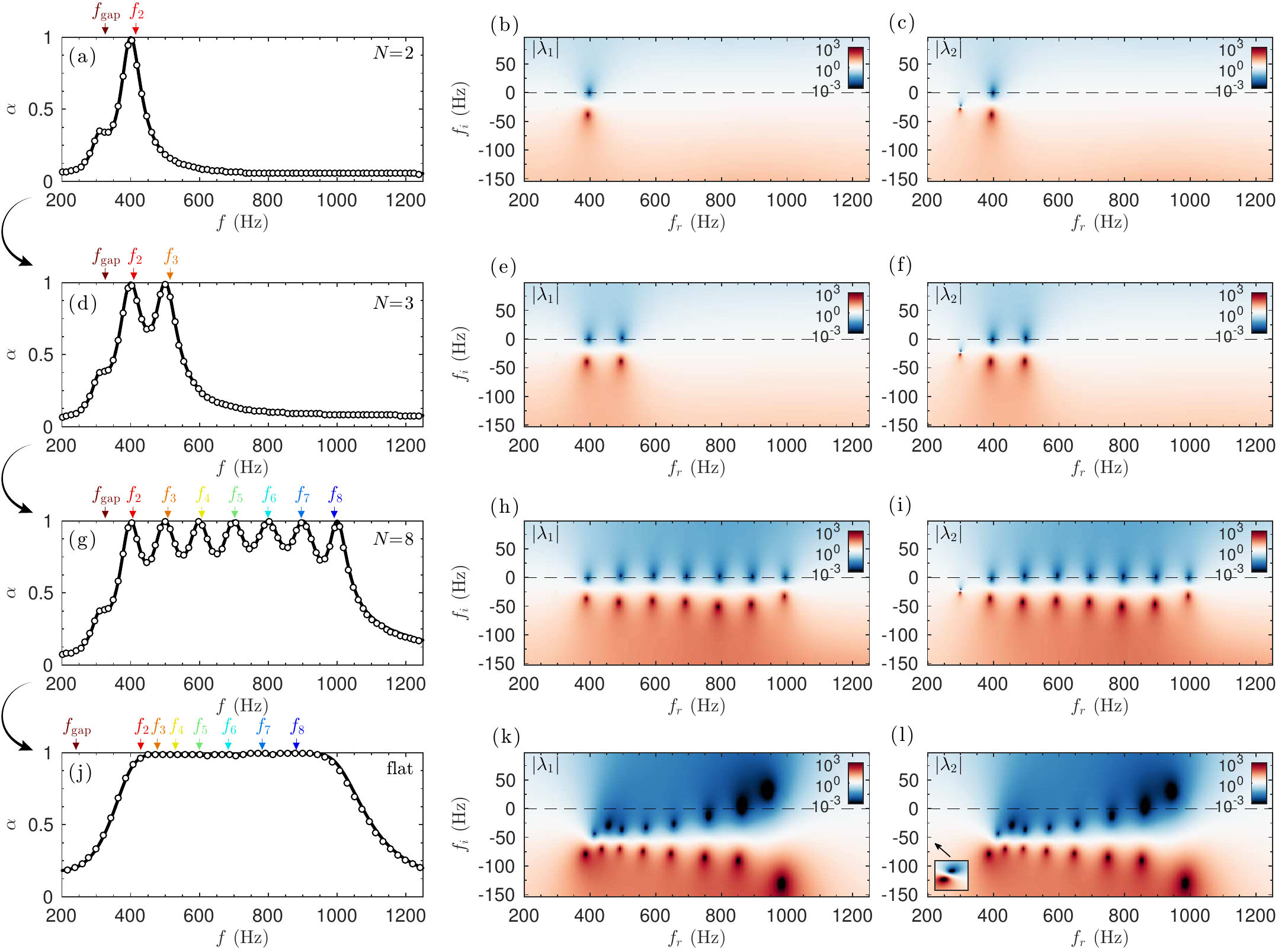}
	\caption{Process to critically couple the rainbow-trapping absorbers. (a) Absorption using $N=2$ HRs obtained using TMM (continuous lines) and FEM simulations (markers), (b-c) corresponding complex frequency plane representation of the eigenvalues of the scattering matrix. (d) Absorption using $N=3$ HRs and (e-f) corresponding complex frequency plane. (g) Absorption using $N=8$ HRs and (h-i) complex frequency plane. (j) Optimized broadband and flat absorption using $N=8$ HRs, (k-l) corresponding complex frequency plane.}
	\label{fig:Fig4}
\end{figure*}

% ============================================================================
\subsection{Rainbow-trapping absorbers (RTA)}\label{sec:rainbow}
%\subsubsection*{Optimal broadband absorption}\label{sec:optimalrainbow}
The concept of the SAP can be applied to design broadband perfect absorbers. The idea is to create a frequency-cascade of band-gaps and critically coupled resonators in order to generate a rainbow-trapping effect. The process is as follows. First, we tune the deepest resonator $(n=1)$ in the waveguide to reduce the transmission above a frequency $f_{1}$. Second, in the same way as previously done in the SAP, a second resonator with slightly higher resonance frequency, $f_{2}$, is placed in the preceding segment of the waveguide. The geometry of this resonator and the section of the waveguide are tuned to impedance match the system at this frequency. Therefore, the reflection vanishes and a peak of perfect absorption is achieved in the same way as in the SAPs. Note this latter HR also reduces the transmission at even higher frequencies. Then, the process can be repeated by extending the waveguide with more segments, each one with a tuned HR being its resonance frequency higher than the preceding one.

Following this process, a rainbow-trapping absorber (RTA) is designed using $N=8$ resonators. Figure~\ref{fig:Fig4} depicts the design process in detail. First, a panel composed of $N=2$ HRs is optimized. Here, the geometry of the metamaterial is tuned in the same way as in the SAP. Figure~\ref{fig:Fig4}~(a) shows a peak of perfect absorption and, as shown in Figs.~\ref{fig:Fig4}~(b-c), both eigenvalues of the scattering matrix present a zero at the same real frequency (as we have already discussed the relevance of the eigenvectors in the previous Section, here we only show the eigenvalues for the sake of simplicity). Then, another HR is added with a slightly higher resonance frequency. The system is again tuned, but this time looking for perfect absorption at two single frequencies, 300 and 320 Hz. Figure~\ref{fig:Fig4}~(d) shows the obtained absorption coefficients where two peaks of perfect absorption are observed, as demonstrated also by the location of the zeros of the eigenvalues of the scattering matrix on the real frequency axis, shown in Figs.~\ref{fig:Fig4}~(e-f). This process is repeated iteratively until $N=8$ HRs were included. Figures~\ref{fig:Fig4}~(g-i) show the $N-1$ peaks of perfect absorption and the corresponding complex frequency plane representation of the eigenvalues of the scattering matrix, respectively. Note that strictly speaking perfect absorption can only be achieved at $N-1$ singular frequencies due to frequency cascade effect. However, flatter absorption can be generated due to the overlapping of the zeros of $\lambda_{1,2}$ if the quality factor of the resonances is reduced. This can be achieved by tuning again the geometrical parameters of the system, using as the initial condition to the optimization algorithm the previous geometry, and using a cost function covering a broad frequency band as $\varepsilon_{\rm RTA}=\int_{f_1}^{f_N}|R^-|^2+|T|^2 df$. The optimized geometrical parameters are listed in Table \ref{table:tableRTA8}. The final flatter, broadband and quasi-perfect absorption is shown in Fig.~\ref{fig:Fig4}~(j). It is worth noting here that using a wide-bandwidth cost function does not ensure that all the resonances remain critically coupled, as it is shown in the complex frequency plane representation of the eigenvalues of the scattering matrix in Figs.~\ref{fig:Fig4}~(k-l). However, the ripples in the absorption can be strongly reduced and the total absorption of energy in a frequency band can be maximized.  

%\subsubsection*{Experimental validation}\label{sec:experientrainbow}
\begin{figure*}[t]
	\centering
	\includegraphics[width=\textwidth]{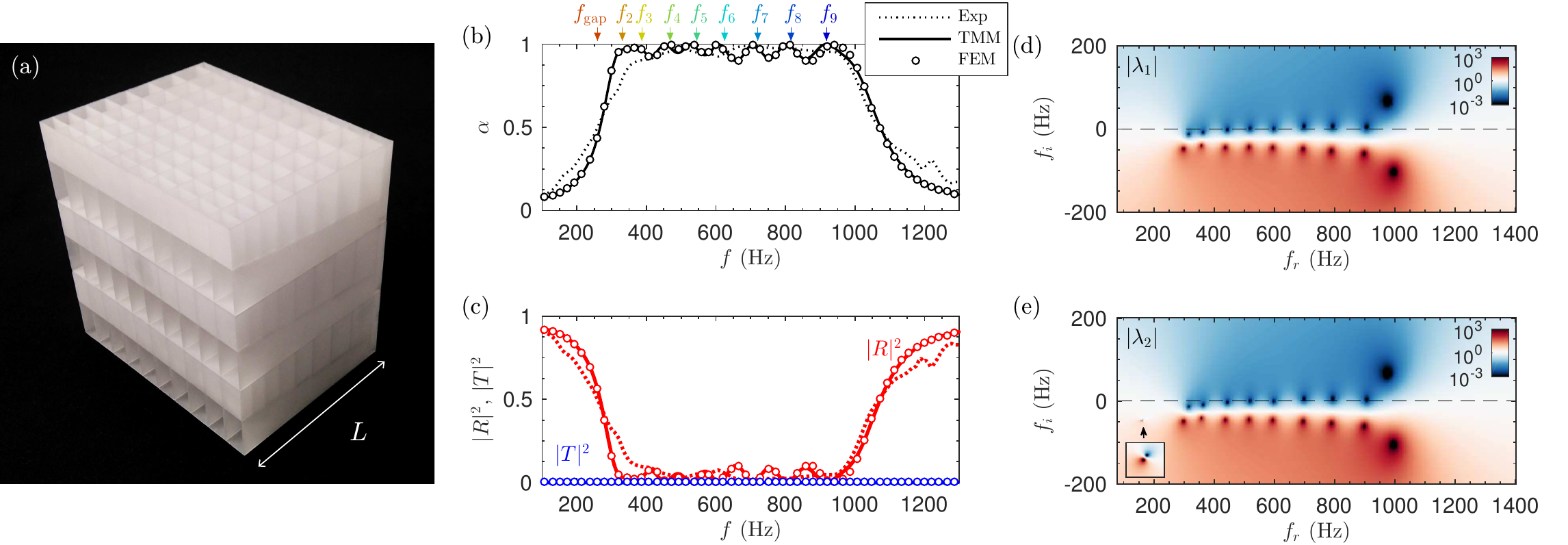}
	\caption{(a) Photograph of the sample containing $10\times3$ unit cells. (b) Absorption obtained by using the TMM (continuous line), FEM simulations (circles) and measured experimentally (dotted line). (b) Corresponding reflection (red curves) and transmission (blue curves).~(d-e) Complex frequency representation of the eigenvalues of the scattering matrix, $\lambda_{1,2}$. Colormap in $10 \log_{10}|\lambda|^2$ scale.}
	\label{fig:Fig5}
\end{figure*}

Due to machine precision of the available 3D printing system (the minimum step was 0.1 mm), the RTA presented previously cannot be easily manufactured. The main limitation was related to the loss of accuracy of the diameters of the small necks that compose the HRs. Under this technological constraint we redesign the RTA using $N=9$ HRs and quantizing the dimensions of all the geometrical elements that compose the structure to the machine precision. The manufactured sample is shown in Fig.~\ref{fig:Fig5}~(a) and the quantized geometrical parameters are listed in Table~\ref{table:tableRTA9}. Figures~\ref{fig:Fig5}~(b-c) show the absorption, reflection and transmission of the device calculated with the TMM, FEM and measured experimentally. The deepest resonator $(n=1)$ presents a resonance frequency of $f_\mathrm{1}=f_\mathrm{gap}=259$ Hz, causing the transmission to drop. A set of 8 resonators were tuned following the process previously described, with increasing resonance frequencies ranging from 330 to 917 Hz. As a result of the frequency-cascade process, the impedance of the structure in the working frequency range is matched with the exterior medium while the transmission vanishes. As a consequence, the RTA presents a flat and quasi-perfect absorption coefficient in this frequency range (see Fig.~\ref{fig:Fig5}~(b)). Excellent agreement is found between the TMM predictions and FEM simulations, while good agreement is observed between the experimental measurements and both models. It can be observed that at low frequencies there are small differences between the measurements and the models. These disagreements are mainly caused by imperfections in the sample manufacturing, by imperfect fitting of the structure to the impedance tube, by the possible evanescent coupling between adjacent waveguides and adjacent HRs, and/or by the limitations of the visco-thermal model used at the joints between waveguide sections. 

\begin{figure*}[t]
	\centering
	\includegraphics[width=0.8\textwidth]{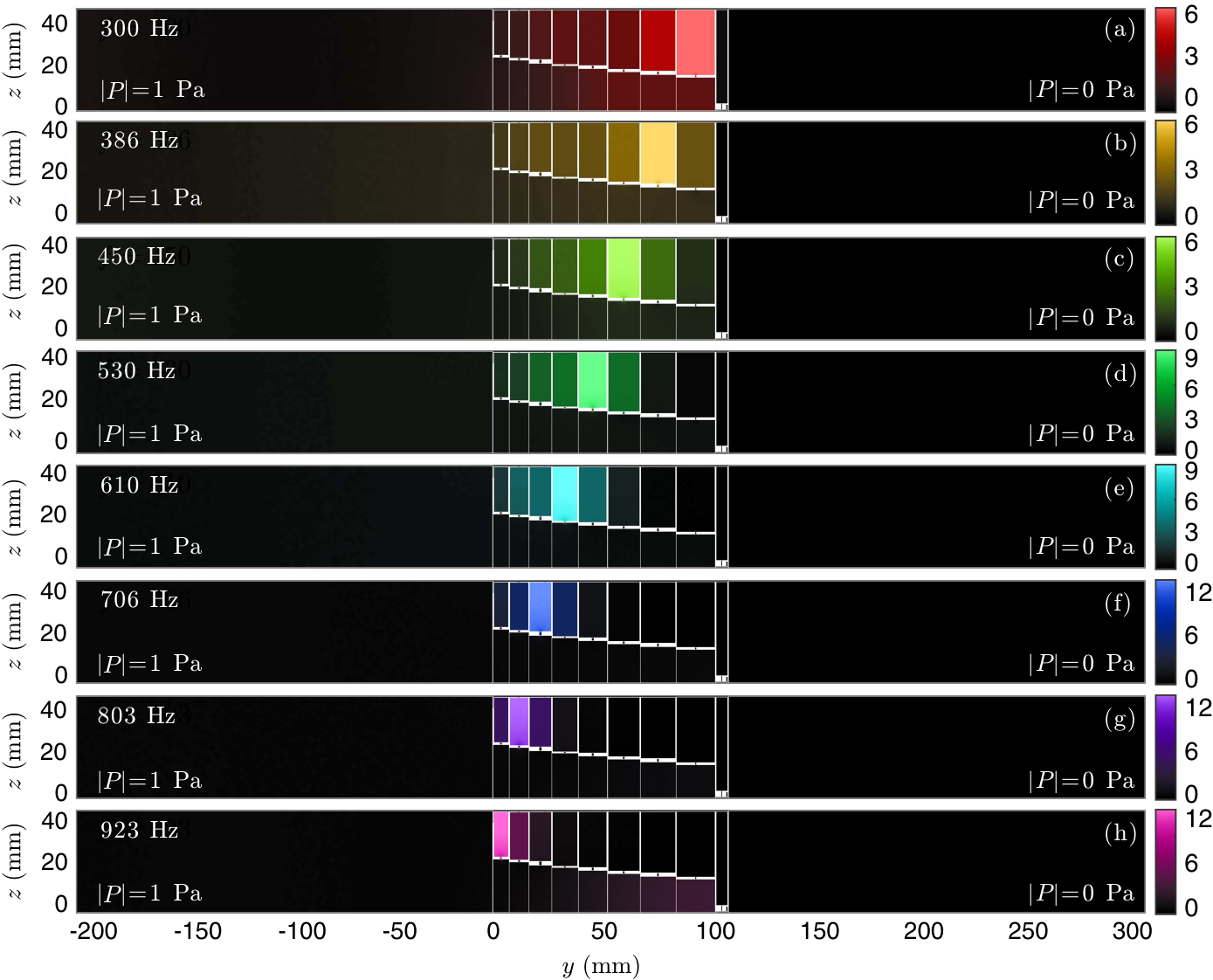}
	\caption{(a-h) Pressure field inside the structure at $x_1=d_1/2$ for frequencies corresponding to the peaks of absorption, i.e., $f=[300, 386, 450, 530, 610, 706,803, 923]$ Hz. Colormap in $|P/P_0|^2$ units, and $P_0=1$ Pa.}
	\label{fig:Fig6}
\end{figure*}

The corresponding representation of the the two eigenvalues of the \textbf{S}-matrix in the complex frequency plane is shown in Figs.~\ref{fig:Fig5}~(d-e). We can see that even under the constraints imposed by the metamaterial construction process, all the $N-1$ zeros of the eigenvalues that produce the critical coupling of the structure are located very close to the real axis being the zeros of $\lambda_1$ at the same frequencies as $\lambda_2$. Note in the manufactured system, not all the zeros are located exactly on the real axis, but the quality factor of the resonances is very low (note the logarithmic colour scale in Fig.~\ref{fig:Fig5}~(c-d)). Therefore they overlap producing quasi-perfect sound absorption in a frequency band from 300 to 1000 Hz for a panel 10 times thinner than the wavelength at 300 Hz in air.

Finally, the pressure field calculated using FEM simulations is shown in Fig.~\ref{fig:Fig6} along the sagittal plane $x_1=d_1/2$ for frequencies corresponding to the peaks of absorption, i.e., $f=[300, 386, 450, 530, 610, 706,803, 923]$ Hz. It can be observed that for each frequency the acoustic field is mostly localized in a single resonator, being the field at lower frequencies localized at the deepest resonator, and the higher frequencies at the outer HRs of the metamaterial, thus, creating the rainbow-trapping effect. Note this behaviour is somehow similar to what was observed in sawtooth broadband absorbers \cite{cui2012,ding2012,jiang2014}, but here transmission was considered. It is worth noting here that for low number of resonators, e.g., the previous cases of $N=2$ and $N=3$ in Figs.~\ref{fig:Fig4}~(a,d), the change of section was not mandatory. However, to obtain broadband absorption, the optimization of the geometry produces always a graded profile of the main waveguide. This graded profile helps to obtain the broadband impedance matching by making use of the cavity resonance in the main waveguide. This cavity mode can be observed at around 923 Hz in the main waveguide, see Fig.~\ref{fig:Fig6}~(h), and is produced by the quarter-wavelength resonance of the waveguide $f \approx c/4\sum_{1}^{N-1}a^{[n]}$. We notice that this resonance was also observed at the complex frequency plane in Fig.~\ref{fig:Fig5}~(d-e), represented by the pairs of zero-poles located at $f\approx 1000 \pm i 100$ Hz. Due to the geometrical constraints of the metamaterial, the critical coupling of this quarter-wavelength resonance is not possible. However, this cavity resonance also contributes, in a moderate way to generate broadband and flat absorption. It is worth noting here that a similar cavity resonance was also visible in Figs.~\ref{fig:Fig4}~(k-l). Therefore, the graded profile of the main waveguides contributes also to generate the flat absorption curve.

% ============================================================================
\section{Discussion}\label{sec:Conclusions}
We reported perfect acoustic absorption over a broad frequency band in deep-subwavelength thickness panels including transmission using the rainbow-trapping effect. In particular, we first presented monochromatic perfect absorption for a subwavelength asymmetric panel (SAP) 40 times smaller than the incoming wavelength (2.64 cm at 300 Hz) using two double-interacting Helmholtz resonators. Then, we reported flat and perfect absorption over a frequency range covering from 300 to 1000 Hz, i.e., almost two octaves, using a rainbow-trapping absorber (RTA) composed of nine resonators and ten times smaller than the wavelength at 300 Hz (11.3 cm). We showed that to obtain broadband and perfect absorption in the transmission problem, three conditions must be simultaneously fulfilled: (\textit{i}) the zeros of the eigenvalues of the scattering matrix must be located on the real frequency axis, (\textit{ii}) the zeros of both eigenvalues, $\lambda_{1,2}$, must be at same frequencies, and, (\textit{iii}) the quality factor of the resonances must be low to overlap in frequency. These three conditions are mandatory to maximize the broadband absorption of the panels and were satisfied by the optimization process.

The limitations to obtain perfect absorption in realistic panels were tested. It is
worth noting here that although achieving perfect absorption is theoretically possible, the maximum absorption is, in general, limited in a real situation. Factors as the changes in temperature, the contribution of nonlinearity for finite amplitude waves or manufacturing constraints produce small changes in the resonances and the critical coupling conditions cannot be exactly fulfilled. However, under such considerations we demonstrated that the structures analysed here produce extraordinary values of absorption: for SAP an absorption peak of 0.995 in the analytical model and 0.982 in the measurements was observed, while in the RTA the maximum absorption was 0.999 in the analytical model and 0.989 in the measurements.

The metamaterials presented  here paves the way to new investigations by using the rainbow-trapping effect produced by other types of resonators as graded arrangements of membranes or poroelastic plates. The current subwavelength thickness configuration using Helmholtz resonators can also has potential applications managing acoustic waves in civil, automotive or aerospace engineering.

%\bibliographystyle{aipauth4-1}
%\bibliography{HRtransmission}

\section{Methods}
\subsection{Theoretical model}
The theoretical modelling is performed by using the transfer matrix method (TMM), which relates the sound pressure, $p$, and normal acoustic particle velocity, $v_x$, at the beginning ($x=0$) and at the end of the panel $(x=L)$. Under the assumption that only plane waves propagate in the waveguides, the transfer matrix ${\bf{T}}$ is derived and the reflection and transmission coefficients can be calculated. The system is written as
\begin{equation}
\left[\begin{tabular}{c}
$p$\\
$v_x$
\end{tabular}\right]_{x=0}= {\bf{T}} \left[\begin{tabular}{c}
$p$\\
$v_x$
\end{tabular}\right]_{x=L} =
\left[\begin{tabular}{cc}
$T_{11}$ & $T_{12}$\\
$T_{21}$ & $T_{22}$
\end{tabular}\right]
\left[\begin{tabular}{c}
$p$\\
$v_x$
\end{tabular}\right]_{x=L},
\end{equation}
\noindent where  ${\bf{T}}$ is given by the product of the transfer matrices of the $N$ different cross-section waveguides loaded by HRs,
\begin{equation}\label{eq:totalmatrix}
{\bf{T}} =  {\bf M}^{[1]}_{\Delta l_{\mathrm{slit}}} \prod_{n=1}^{N} {\bf M}_{\bf s}^{[n]} {\bf M}_\mathrm{HR}^{{[n]}}{\bf M}_{\bf s}^{[n]} {\bf M}_{\Delta l_{\mathrm{slit}}}^{[n+1]}\,.
\end{equation}
The transmission matrix for the $n$-th waveguide half-segment, ${\bf M}_{\bf s}^{[n]}$, takes the form 
\renewcommand{\arraystretch}{1.5}
\begin{align}
{\bf M}_{\bf s}^{[n]}=\left[ 
\begin{array}{cc} 
\cos \left(k_{\bf s}^{[n]} \frac{a^{[n]}}{2} \right) & \i Z_{\bf s}^{[n]} \sin \left( k_{\bf s}^{[n]} \frac{a^{[n]}}{2}\right)  \\ 
\frac{\i}{Z_{\bf s}^{[n]}}\sin \left(k_{\bf s}^{[n]} \frac{a^{[n]}}{2}\right)  & \cos  \left(k_{\bf s}^{[n]} \frac{a^{[n]}}{2} \right) 
\end{array} \right],
\end{align}
\noindent where $k_{\bf s}^{[n]}$ and $Z_{\bf s}^{[n]}=\sqrt{\kappa_{\bf s}^{[n]} \rho_{\bf s}^{[n]}}/S_{\bf s}^{[n]}$ are the effective wavenumber and the characteristic impedance of the $n$-th waveguide segment, $\kappa_{\bf s}^{[n]}$ and $\rho_{\bf s}^{[n]}$ are the effective bulk modulus and the density respectively, provided in the \ref{sec:appendix1}, and $S_{\bf s}^{[n]}=d_1^{[n]}\,h_3^{[n]}$ are the cross-sectional areas. The resonators, accounted for as point scatterers in the middle of each waveguide segment by a transmission matrix ${{\bf M}_{\mathrm{HR}}^{[n]}}$, and the radiation correction of the $n$-th waveguide segment due to cross-section changes, ${\bf M}_{\Delta l_{\mathrm{slit}}}^{[n]}$, are respectively
\renewcommand{\arraystretch}{1}
\begin{align}
{{\bf M}_{\mathrm{HR}}^{[n]}}=
\left[ \begin{array}{cc} 
1 & 0 \\ 
1/{{Z}_{\mathrm{HR}}^{[n]}} & 1 
\end{array} \right], \quad
{\bf M}_{\Delta l_{\mathrm{slit}}}^{[n]}=
\left[ \begin{array}{cc} 
1 & {Z}_{\Delta l_{\mathrm{slit}}}^{[n]} \\ 
0 & 1 
\end{array} \right],
\end{align}
\noindent where ${{Z}_{\mathrm{HR}}^{[n]}}$ is the impedance of the HR and ${Z}_{\Delta l_{\mathrm{slit}}}^{[n]}$ is the characteristic radiation impedance of the $n$-th waveguide, both provided in the \ref{sec:appendix2}. Notice that the length corrections of the HRs are already accounted for in the impedance ${{Z}_{\mathrm{HR}}^{[n]}}$.

The reflection coefficients from both sides of the structure, $R^+$ and $R^-$, and the transmission coefficient, $T$, are given by the elements of the \textbf{T}-matrix as
\begin{align}
\label{eq:T}
T=\frac{2e^{\imath k L}}{T_{11}+T_{12}/Z_0+Z_0T_{21}+T_{22}},\\%\quad
%\label{eq:Rm}
R^-=\frac{T_{11}+T_{12}/Z_0-Z_0T_{21}-T_{22}}{T_{11}+T_{12}/Z_0+Z_0T_{21}+T_{22}},\\%\quad%~~,~~
%\label{eq:Rp}
R^+=\frac{-T_{11}+T_{12}/Z_0-Z_0T_{21}+T_{22}}{T_{11}+T_{12}/Z_0+Z_0T_{21}+T_{22}},
\end{align}
\noindent where $Z_0=\rho_0 c_0/S_0$ is the characteristic impedance of the surrounding medium, usually air, with $S_0=d_1 d_3$, and the superscripts ($+,-$) denoting the incidence direction, i.e., the positive and negative $x_2$-axis respectively. Finally the asymmetric absorption coefficients are calculated as $\alpha=\alpha^-=1-\left|R^-\right|^2-\left|T\right|^2$ for the positive $x_2$-axis ingoing waves, namely here and beyond \textit{forward} propagation, and $\alpha^+=1-\left|R^+\right|^2-\left|T\right|^2$ for the negative $x_2$-axis ingoing waves, namely \textit{backward} propagation. In symmetric systems $T_{11}=T_{22}$ and, as a consequence, $R^+=R^-$. This property is not satisfied by rainbow trapping absorbers and, therefore, the absorption depends on the direction of incidence. The reciprocal behaviour of the system implies that the determinant of transfer matrix is one ($T_{11}T_{22}-T_{12}T_{21}=1$). This property is satisfied by the present linear and time invariant system and the transmission does not depend on the direction of incidence.

On the other hand, the scattering matrix, $\bf{S}$, relates the amplitudes of the incoming waves with those of the outgoing waves. The total pressures at both sides of the structure are given by $p(x_a)=Ae^{-\imath k x_a}+Be^{\imath k x_a}$ for $x_a<0$, and $p(x_b)=Ce^{-\imath k x_b}+De^{\imath k x_b}$ for $x_b>L$. Thus, the relation between the amplitudes of both waves is given by the $\bf{S}$-matrix as
\begin{align}
\label{eq:S}
\left[\begin{tabular}{c}
$A$\\
$D$
\end{tabular}\right]=
\bf{S}
\left[\begin{tabular}{c}
$C$\\
$B$
\end{tabular}\right]
=
\left[\begin{tabular}{cc}
$T$ & $R^-$\\
$R^+$ & $T$
\end{tabular}\right]
\left[\begin{tabular}{c}
$C$\\
$B$
\end{tabular}\right].
\end{align}
The $\bf{S}$-matrix is widely used in wave physics to characterize and interpret the wave scattering. The scattering matrix possesses two eigenvalues
\begin{equation}
\lambda_{1,2}=T\mp\sqrt{R^+R^-},\label{eq:lambda}
\end{equation}
while the eigenvectors corresponding to $\lambda_1$ and $\lambda_2$ are 
\begin{align}
\vec{v}_1= \left[R^-,-\sqrt{R^+R^-}\right],\label{eq:v1} \quad
\vec{v}_2=\left[\sqrt{R^+R^-},R^+\right],%\label{eq:v2}
\end{align} 
respectively. The poles and zeros of the eigenvalues as well as the eigenvectors of the $\bf{S}$-matrix in the complex-frequency plane provide rich information, as we will see later.

\subsection{Numerical model}
In order to validate the analytical models we use a numerical approach based on the Finite Element Method (FEM) using COMSOL Multiphysics 5.2\textsuperscript {TM}. The thermo-viscous losses were accounted for by using the effective parameters of the air in the ducts, i.e., by using the complex and frequency dependent density and bulk modulus, given in the \ref{sec:appendix1}. At the external sides of the panel, rigid boundary conditions were considered and viscous losses were neglected here. This is justified because losses are mainly produced by thermo-viscous processes at the narrow ducts that compose the metamaterial and the contribution of other sources is minor. The unstructured mesh was designed ensuring a maximum element size $20$ times smaller than the wavelength.

\subsection{Sample design and experiments}
Both samples were 3D printed by means of stereo-lithography techniques using a photosensitive epoxy polymer (Accura 60{\textsuperscript{\textregistered}}, 3D Systems Corporation, Rock Hill, SC 29730, USA). The acoustic properties of the solid phase were $\rho_0=1210$ kg/m$^3$ and $c_0= 1630 \pm 60$ m/s. Therefore, the characteristic acoustic impedance was almost 5 thousand times greater than the one of air and therefore the structure is considered motionless. The transmission, reflection and absorption were measured in a impedance tube with squared cross-section whose side was 15 cm. During the experiments the amplitude of the acoustic source was low enough to neglect the contribution of the nonlinearity of the HRs.

\subsection{Visco-thermal losses model}\label{sec:appendix1}
The visco-thermal losses in the system are considered both in the HRs and in the waveguide by using its effective complex and frequency dependent parameters. Considering only plane waves propagate inside the metamaterial, the effective density, $\rho_i^{[n]}$, and bulk modulus, $\kappa_i^{[n]}$, of the $n$-th waveguide segment, $\rho_{\bf s}^{[n]}$, $\kappa_{\bf s}^{[n]}$, the neck, $\rho_{\bf n}^{[n]}$, $\kappa_{\bf n}^{[n]}$, and the cavity, $\rho_{\bf c}^{[n]}$, $\kappa_{\bf c}^{[n]}$, of each resonator are given by \cite{stinson1991}:
\begin{align}\label{eq:rhoc}
\rho_i^{[n]} = -\frac{\rho_0 b_1^2 b_2^2}{4 G_\rho^2 \sum\limits_{k\in\mathbb{N}}\sum\limits_{m\in\mathbb{N}} \left[\alpha_k^2 \beta_m^2 \(\alpha_k^2 + \beta_m^2 - G_\rho^2\) \right]^{-1}} \,, \\%\quad
\kappa_i^{[n]} =\frac{\kappa_0}{\gamma + \frac{4 (\gamma -1) G_\kappa^2 }{{b_1^2}{b_2^2}}\sum\limits_{k\in\mathbb{N}}\sum\limits_{m\in\mathbb{N}}{\left[\alpha _k^2\beta _m^2\(\alpha _k^2+\beta _m^2-G_\kappa^2\)\right]^{-1}}} \,, %\label{eq:Kc}
\end{align}

\noindent where $G_\rho=\sqrt{{\i\omega\rho_0}/{\eta}}$ and $G_\kappa=\sqrt{\i\omega\mathrm{Pr}\rho_0/{\eta}}$, $\gamma$ is the specific heat ratio of air, $P_0$ is the atmospheric pressure, $\Pr$ is the Prandtl number, $\eta$ the dynamic viscosity, $\rho_0$ the air density, $\kappa_0={\gamma P_0}$ the air bulk modulus and $\omega$ the angular frequency. The constants $\alpha _k=2(k+1/2)\pi/b_1$ and $\beta_m=2(m+1/2)\pi/b_2$, and $b_1$ and $b_2$ are the dimensions of the duct. Thus, in the case of the $n$-th waveguide segment $b_1= h_1^{[n]}$, $b_2=h_3^{[n]}$; in the case of the neck of the HR $b_1=b_2=w_{\bf n}^{[n]}$; and in the case of the cavity of the HR $b_1=w_{{\bf c},1}^{[n]}$ and $b_2=w_{{\bf c},2}^{[n]}$. Finally, the effective wave number and acoustic impedance are given by $k_i^{[n]}=\omega \sqrt{\rho_i^{[n]}/\kappa_i^{[n]}}$ and $Z_i^{[n]} = \sqrt{\kappa_i^{[n]} \rho_i^{[n]}}/b_1 b_2$ respectively.

\subsection{Resonator impedance and end corrections}\label{sec:appendix2}
Using the effective parameters for the neck and cavity elements given by Eqs.~(\ref{eq:rhoc}-\ref{eq:Kc}), the impedance of a Helmholtz resonator, including a length correction due to the radiation can be written as \cite{theocharis2014}:
\begin{widetext}
\begin{align}
Z_\mathrm{HR}^{n} = -\i \frac{\cos(k_{\bf n}^n l_{\bf n}^n) \cos(k_{\bf c}^n l_{\bf c}^n) - Z_{\bf n}^n k_{\bf n}^n \Delta l^n \cos(k_{\bf n}^n l_{\bf n}^n) \sin(k_{\bf c}^n l_{\bf c}^n)/Z_{\bf c}^n - Z_{\bf n}^n \sin(k_{\bf n}^n l_{\bf n}^n)\sin(k_{\bf c}^n l_{\bf c}^n)/Z_{\bf c}^n}{\sin(k_{\bf n}^n l_{\bf n}^n)\cos(k_{\bf c}^n l_{\bf c}^n)/Z_{\bf n}^n - k_{\bf n}^n \Delta l^n\sin(k_{\bf n}^n l_{\bf n}^n)\sin(k_{\bf c}^n l_{\bf c}^n)/Z_{\bf c}^n + \cos(k_{\bf n}^n l_{\bf n}^n)\sin(k_{\bf c}^n l_{\bf c}^n)/Z_{\bf c}^n}\,,
\end{align}
\end{widetext}
\noindent where $l_{\bf n}^{n}$ and $l_{\bf c}^{n}$ are the neck and cavity lengths, $k_{\bf n}^{n}$ and $k_{\bf c}^{n}$, are the effective wavenumbers and and $Z_{\bf n}^{n}$ and $Z_{\bf c}^{n}$ effective characteristic impedance in the neck and cavities respectively, and $\Delta l^{n}$ the correction length for the HRs. These correction lengths are deduced from the addition of two correction lengths $\Delta l^{n}=\Delta l_1^{[n]} + \Delta l_2^{[n]}$ as
\begin{align}
\Delta l_1^{[n]} = &0.41 \left[1 - 1.35 \frac{w_{\bf n}^{[n]}}{w_{\bf c}^{[n]}} + 0.31 \(\frac{w_{\bf n}^{[n]}}{w_{\bf c}^{[n]}}\)^3\right] w_{\bf n}^{[n]} \,,\\
\Delta l_2^{[n]} = &0.41 \left[1 - 0.235 \frac{w_{\bf n}^{[n]}}{w_{\bf s}} - 1.32\(\frac{w_{\bf n}^{[n]}}{w_{\bf s}}\)^2 \nonumber\right.\\&\left.+ 1.54 \(\frac{w_{\bf n}^{[n]}}{w_{\bf s}}\)^3 - 0.86\(\frac{w_{\bf n}^{[n]}}{w_{\bf s}}\)^4 \right] w_{\bf n}^{[n]}\,. \nonumber
\end{align}

The first length correction, $\Delta l_1^{[n]}$, is due to pressure radiation at the discontinuity from the neck duct to the cavity of the HR \cite{kergomard1987}, while the second $\Delta l_2^{[n]}$ comes from the radiation at the discontinuity from the neck to the principal waveguide \cite{dubos1999}. This correction only depends on the radius of the waveguides, so it becomes important when the duct length is comparable to the radius.

Another important end correction arises due the radiation from the waveguides to the free air, and the radiation correction between waveguides segments due to change of section. On one hand, in the case of radiation to the free air, for the element $n=N$ in Eq.(\ref{eq:totalmatrix}), the radiation correction is equivalent to the one of a periodic distribution of slits, that can be expressed as \cite{mechel2013}: 
\begin{equation}\label{eq:Dslit}
\Delta l_{\rm slit}^{[N]} = \sigma h_3^{[N]} \sum_{m=1}^{\infty}\frac{\sin^2\(m\pi \sigma\)}{(m\pi \sigma)^3},
\end{equation}
\noindent with $\sigma = h_3^{[N]}/d_3$. Using these values, the radiation impedance of the $N$ waveguide segment is ${Z}_{\Delta l_{\mathrm{slit}}}^{[N]}=-i\omega\Delta l_\mathrm{slit}^{[N]}\rho_0/\phi_t^{[N]} S_0$ with $\phi^{[N]}_t = d_3/h_3^{[N]}$ and the unit cell surface $S_0=d_1d_3$. On the other hand, for the radiation correction between slits due to change of section, i.e., from $n=1$ to $n=N-1$ in Eq.(\ref{eq:totalmatrix}), the following end correction has been applied:
\begin{equation}\label{eq:Dslit2}
\Delta l_{\rm slit}^{[n]} = 0.82\left[ 1- 1.35 {\frac{h^{[n]}}{h^{[n-1]}}} + 0.31 \left( \frac{h^{[n]}}{h^{[n-1]}}\right)^3 \right] h^{[n]}.
\end{equation}
\noindent Using this value, the radiation impedance reads ${Z}_{\Delta l_{\mathrm{slit}}}^{[n]}=-i\omega\Delta l_\mathrm{slit}^{[n]}\rho_0/\phi_t^{[n]} S_{\bf s}^{[n]}$ with $\phi^{[n]}_t = h^{[n-1]}/h^{[n]}$ and $S_{\bf s}^{[n]}=h_1^{[n]} h_3^{[n]}$.

\subsection{Geometrical parameters}\label{appendix:c}
The geometrical parameters for the SAP $(N=2)$, corresponding to Fig.~\ref{fig:Fig2}, are listed in Table \ref{table:tableSAP2}. The total structure thickness is $L=\sum a^{[n]} = 28.6$ mm, and its height and width of the unit cell are  $d_3=148.1$ mm and $d_1 = 14.8$ mm respectively.
\begin{table*}[ht]                
	\centering \footnotesize                  
	\begin{tabular}{lcccccccc}            
		\hline               
		$n$ & $a^{[n]}$ (mm) & $h_3^{[n]}$ (mm) & $h_1^{[n]}$ (mm) & $l_{\bf n}^{[n]}$ (mm) & $l_{\bf c}^{[n]}$ (mm) & $w_{\bf n}^{[n]}$ (mm) & $w_{{\bf c},1}^{[n]}$ (mm) & $w_{{\bf c},2}^{[n]}$ (mm) \\
		\hline
		2 & 16.8 & 12.7 & 13.8 & 15.4 & 119.1 & 4.5 & 13.8 & 15.7 \\                                                                                      
		1 & 11.8 & 1.0 & 13.8 & 12.0 & 134.1 & 3.2 & 13.8 & 10.8 \\ 
		\hline                                                                                      
	\end{tabular}                                                                                                                                       
	\caption{Geometrical parameters for the SAP $(N=2)$.}      
	\label{table:tableSAP2}    
\end{table*}    

The geometrical parameters for the RTA $(N=8)$, corresponding to Fig.~\ref{fig:Fig4}~(j-l), are listed in Table \ref{table:tableRTA9}. The total structure thickness is $L = 120$ mm, and its height and width of the unit cell are  $d_3=48.7$ mm and $d_1 = 6.3$ mm respectively.
\begin{table*}[ht]                                                                                                                                      
	\centering  \footnotesize                                                                                                                                             
	\begin{tabular}{lcccccccc}                                                                                                                      
		\hline    
		$n$ & $a^{[n]}$ (mm) & $h_3^{[n]}$ (mm) & $h_1^{[n]}$ (mm) & $l_{\bf n}^{[n]}$ (mm) & $l_{\bf c}^{[n]}$ (mm) & $w_{\bf n}^{[n]}$ (mm) & $w_{{\bf c},1}^{[n]}$ (mm) & $w_{{\bf c},2}^{[n]}$ (mm) \\
		\hline
		8 & 15.0 & 28.8 & 5.3 & 1.0 & 17.9 & 0.8 & 5.3 & 13.5 \\                                                                                          
		7 & 15.0 & 27.4 & 5.3 & 1.2 & 19.2 & 0.8 & 5.3 & 14.0 \\                                                                                          
		6 & 15.0 & 25.9 & 5.3 & 1.5 & 20.3 & 0.8 & 5.3 & 14.0 \\                                                                                          
		5 & 15.0 & 24.5 & 5.3 & 1.8 & 21.3 & 0.8 & 5.3 & 14.0 \\                                                                                          
		4 & 15.0 & 23.1 & 5.3 & 2.2 & 22.4 & 0.7 & 5.3 & 14.0 \\                                                                                          
		3 & 15.0 & 22.1 & 5.3 & 2.6 & 23.0 & 0.7 & 5.3 & 14.0 \\                                                                                          
		2 & 15.0 & 21.9 & 5.3 & 3.2 & 22.6 & 0.7 & 5.3 & 14.0 \\                                                                                          
		1 & 15.0 & 0.8 & 0.8 & 5.7 & 41.2 & 0.7 & 5.3 & 13.5 \\  
		\hline                                                                                         
	\end{tabular}                                                                                                                                       
	\caption{Geometrical parameters for the RTA $(N=8)$.}                                                                                                                                 
	\label{table:tableRTA8}                                                                                                                          
\end{table*}     

The geometrical parameters for the RTA $(N=9)$, corresponding to Fig.~\ref{fig:Fig5}, are listed in Table \ref{table:tableRTA9}. The total structure thickness is $L=\sum a^{[n]} = 113$ mm, and its height and width of the unit cell are  $d_3=48.7$ mm and $d_1 = 14.6$ mm respectively.
\begin{table*}[ht]                                                                                     
	\centering  \footnotesize                                                                                     
	\begin{tabular}{lcccccccc}     
		\hline                                                                      
		$n$ & $a^{[n]}$ (mm) & $h_3^{[n]}$ (mm) & $h_1^{[n]}$ (mm) & $l_{\bf n}^{[n]}$ (mm) & $l_{\bf c}^{[n]}$ (mm) & $w_{\bf n}^{[n]}$ (mm) & $w_{{\bf c},1}^{[n]}$ (mm) & $w_{{\bf c},2}^{[n]}$ (mm) \\
		\hline
		9 & 7.9 & 25.6 & 14.0 & 1.1 & 21.4 & 1.2 & 14.0 & 7.2 \\                                              
		8 & 9.5 & 24.2 & 14.0 & 1.0 & 22.8 & 1.2 & 14.0 & 9.0 \\                                              
		7 & 11.0 & 22.8 & 14.0 & 1.7 & 23.6 & 1.4 & 14.0 & 10.6 \\                                            
		6 & 12.6 & 21.6 & 14.0 & 0.7 & 25.9 & 1.0 & 14.0 & 12.0 \\                                            
		5 & 14.1 & 20.2 & 14.0 & 1.5 & 26.5 & 1.2 & 14.0 & 13.6 \\                                            
		4 & 15.7 & 18.8 & 14.0 & 1.1 & 28.3 & 1.0 & 14.0 & 15.2 \\                                            
		3 & 17.3 & 17.4 & 14.0 & 1.6 & 29.2 & 1.0 & 14.0 & 16.8 \\                                            
		2 & 18.8 & 16.0 & 14.0 & 1.1 & 31.2 & 0.8 & 14.0 & 18.4 \\                                            
		1 & 6.4 & 1.0 & 1.0 & 3.0 & 44.7 & 0.6 & 14.0 & 5.6 \\                                                
		\hline                                                 
	\end{tabular}                                                                                       
	\caption{Geometrical parameters for the RTA $(N=9)$.}                                                                            
	\label{table:tableRTA9}                                                                          
\end{table*}  

\section*{Acknowledgements}
The authors acknowledge financial support from the Metaudible Project No. ANR-13-BS09-0003, cofunded by ANR and FRAE.

\section*{Author contributions statement}
N.J., V.RG. and JP.G. conducted the theoretical modelling and numerical experiment; N.J., V.RG. JP.G. and V.P. wrote the manuscript. All authors reviewed the manuscript.

%\section*{Additional information}
%\textbf{Supplementary information} accompanies this paper at  \href{http://www.nature.com/srep}{http://www.nature.com/srep}\\
%\noindent\textbf{Accession codes} (where applicable); \\
\noindent\textbf{Competing financial interests} The authors declare no competing financial interests.

\end{document}